\documentstyle[nato,numreferences]{crckapb}  

\input{epsf}

\begin{opening} 

\title{THE SHELL MAP} 
\subtitle{The structure of froths through a dynamical map}
 
\author{TOMASO ASTE }

\institute{Laboratoire de Dynamique des Fluides Complexes, \\
Universit\'e Louis Pasteur Strasbourg, 67084 France.\\
tomaso@ldfc.u-strasbg.fr} 
\end{opening} 
 
\runningtitle{THE SHELL MAP}

\begin{document}

\section{\label{i} Introduction} 
\noindent 
The shell map is a very simple representation of the structure of foams,
combining the geometrical (random tiling) and dynamical (loss of information 
from an arbitrary cell out) aspects of disorder. 
We will illustrate it and give several examples, including a few arising from 
discussions in Cargese.
This chapter is written by following the main lines of two previously
published papers \cite{AsBoRi,ShapeM}.

\begin{figure}
\vspace{-6.5cm}
%\hspace{1.5cm}
\epsfxsize=14.cm
\epsffile{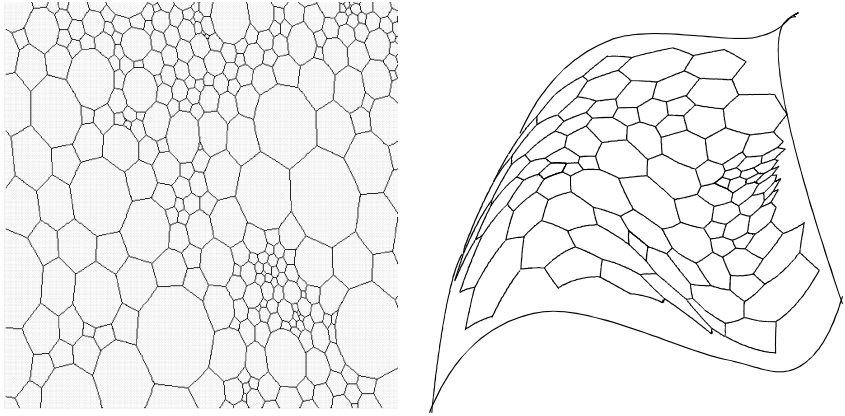} 
\vspace{-7.cm}
\caption{\footnotesize \label{f.froth}\label{f.curv_froth}
An example of two dimensional disordered cellular 
structure (froth) generated by cell division and coalescence transformations (left).
A froth on a curved space (right).}
\vspace{0.cm}
\end{figure}

\begin{figure}
\vspace{-5.5cm}
\epsfxsize=12.cm
%\hspace{2.cm}
\epsffile{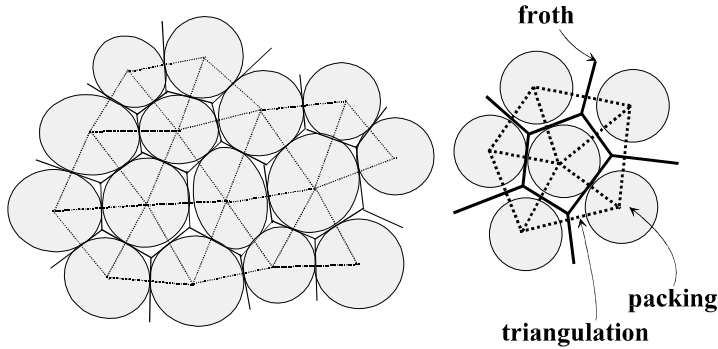}
\vspace{-7.cm}
\caption{\footnotesize \label{f.froth_triang}
Froths and triangulations are dual structures.
Triangulations are useful representations of packings.}
\vspace{0.cm}
\end{figure}

In Nature, space-filling disordered patterns and cellular structures are wide\-spread \cite{dAT,WR84}.
These structures (froths) are partitions of $D$-dimensional space by convex cells. 
Disorder imposes that each vertex has minimal 
number of incident edges, faces and cells ($D+1$ edges incident on a vertex, $D$ faces incident 
on a edge, $D-1$ cells incident on a face, in $D$-dimensions, Fig.\ref{f.froth}). 
In this respect, a froth is a regular graph, but the number of edges bounding each face, 
the number of faces bounding a polyhedral cell, etc., are random variables \cite{AsRi}.  
Minimal incidences implies also that the topological dual of a froth is a triangulation 
(Fig.\ref{f.froth_triang}), a useful representation of packings.  
Indeed, for any given packing, or point set, one can construct the Vorono\"{\i} tessellation
\cite{Voronoi}, which is a space-filling assembly of polyhedral cells.
When the starting points are disordered (no special symmetries) the Vorono\"{\i} tessellation
is a froth.
The space filled by the froth can be curved (Fig.\ref{f.froth}).
This is for instance the case in amphiphilic membranes \cite{CS}, fullerenes, 
the basal layer of the epidermis of mammals \cite{RD} or the ideal structure of amorphous 
materials \cite{SM85,SR}.
Disorder does not necessarily imply inhomogeneity. 
On the contrary, in many cases, disordered froths are very homogeneous (cells with very 
similar sizes and regular shapes). 
This is -for instance- the case in the epidermis, where the biological cells have homogeneous 
sizes and isotropic shapes but the structure is disordered. 
Indeed, the disordered arrangement is the one which best guarantee both the partition of the 
curved space into similar cells and the invariance under mitosis and detachment.

\vspace{0.cm}

The interplay between disorder and curvature is illustrated in this chapter by representing
the froth as organized in concentric layers of cells around an
arbitrary central cell \cite{AsBoRi} (Figs.\ref{f.SSIfroth} and \ref{f.nonSSIfroth}).  
The structure is built from the central cell outward like an ever expanding jigsaw puzzle without boundary.  
The radial map, from one spherical layer of cells to the next, is the
logistic map \cite{Sch84}, and the geometrical tiling is expressed mathematically as a dynamical
system \cite{RiUnp}.
The isotropy of the disordered structure is expressed locally by averaging over each layer.  
The over-all translational invariance is manifest in the independence of the structure 
and properties on the choice of the central cell. 
The radial map from one layer to the next includes both effects of disorder and of space curvature.

In a two dimensional froth, given a cell with $n$ edges, one can define its topological charge as  $q=6-n$.
The total charge of a froth with $N$ cells is the sum over the charges of each cell:
$Q_T = \sum_i q_i = (6- \langle n \rangle)N = 6 \chi$, with $\chi$ the Euler-Poincar\'e characteristic  
of the manifold tiled by the froth \cite{EulerPoincare,G-B}.
The charge is a topological invariant, it cannot be generated or destroyed and the local topological 
transformations in the froth redistribute it between adjacent cells. 
The total charge is equal to zero in Euclidean froths, it is 12 for froths on the surface of a sphere and it is negative
in spaces with negative Gaussian curvature.
In two dimensions, it is therefore possible to define the curvature of a surface tiled by a froth
by analyzing the local topological configuration of its tiles.
In three dimensions, for $N \rightarrow \infty$ or in closed elliptic froths,
 the Euler's relation is homogeneous ($\chi = 0$, independently on the space-curvature) 
and from the Gauss-Bonnet formula \cite{G-B} it is not possible to distinguish the global curvature of a tiled manifold 
from the local average properties of its tiles. 
On the other hand, we show that,  also in three dimensions, the map makes possible to define the curvature of the 
space  from local topological configurations \cite{AsBoRi}.
This is done by simply computing the number of cells in successive layers.

In this chapter, we describe the map which gives the number of cells $K_j$ in a layer distant $j$
from the central cell as function of the average topological properties of the cells in 
the previous layers (section \ref{S.Map}) \cite{AsBoRi}.
We discuss the link between the map and the space curvature giving examples in two and three
dimensions (section \ref{S.MapCurv}) \cite{AsBoRi,ShapeM}.
By using the map we exploit the freedom of constructing a froth with different local topological 
configurations and determine the three dimensional Euclidean structures which maximize such 
freedom (section \ref{S.MapFree}) \cite{AsBoRi,ShapeM}.

\section{\label{s1} From a cell to the whole froth, a topological map} \label{S.Map}
\begin{figure}
\vspace{-1.5cm}
%\hspace{2.cm}
\epsfxsize=8.cm
\epsffile{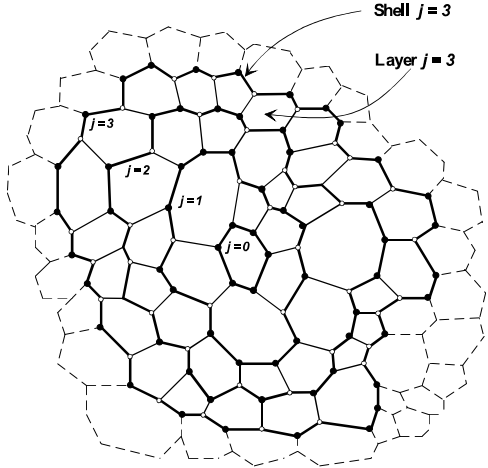}
\vspace{-4.cm}
\caption{\footnotesize \label{f.SSIfroth}
An example of SSI froth. 
The number of cells in each layer can be calculated in term of a simple map }
\vspace{0.cm}
\end{figure}

\noindent
All froths can be studied as structured in concentric layers of cells 
which are at the same topological distance from a given central cell.
The topological distance between two cells is the minimum number of edges crossed by a path from
one cell to the other \cite{AsBoRi}.
The layers are closed rings of irregular polygons in two dimensions and spherical caps of 
irregular polyhedra in three dimensions.
The cells making the layer $j$ can be distinguished in two categories. 
Some cells have simultaneously neighbours in the layers $j-1$ and $j+1$, these cells make themselves 
closed layers and constitute the {\it skeleton } of the shell-structure.
Other cells (or clusters of cells) are local inclusions ({\it topological defects}) between the layers of the 
shell-skeleton (they don't have neighbours in the layer $j+1$).
The shell-skeleton is itself a space-filling froth, hierarchically organized around the germ cell.
Once the germ cell is chosen, the shell-structure and its skeleton are univocally defined, 
but different germ cells generate different skeletons.
We call {\it shell-structured-inflatable} (SSI) a froth free of topological defects.
In this case, the shell-structure and its skeleton coincide.
An example of a two dimensional SSI disordered cellular structure is given in Fig.(\ref{f.SSIfroth}). 
A three-dimensional regular SSI structure (the Kelvin froth \cite{Kelvin}) is given in 
Fig.(\ref{f.Kelvin}).
In this paragraph, we first study SSI froths and then generalize the results to generic froths.

\subsection{ A topological map for SSI froths}
\noindent
For SSI structures, a recursive equation gives the number of cells $K_j$ in a layer at a distance 
$j$ from the germ cell, as a function of the number of cells in the previous two layers. 
In two dimensions, one can easely verify that \cite{AsBoRi,RiUnp,TroUnp,Fortes} (see Fig.\ref{f.SSIfroth})
\begin{equation}
K_{j+1} = s_j K_j - K_{j-1} \;\;\;\;\;(\mbox{for $j \ge 1$}),
\label{2Drec}
\end{equation}
where the inflation parameter $s_j$ is related to the average topological properties of the cells in 
layer $j$,
\begin{equation}
s_j = \langle n \rangle_j-4\;\;\;\;\;\;\;\;\;\;,
\label{2Ds}
\end{equation}
with $\langle n \rangle_j$ the average number of neighbours per cell in layer $j$.
The map starts with $K_0 = 0$ and $K_1 = n = $ {\it number of neighbours of the central cell}.
For example in the hexagonal lattice one has $n = 6$, $\langle n \rangle_j =6$ and therefore
$s_j = 2$. 
In this case the map (\ref{2Drec}) correctly gives $K_j = 6 j$.

\begin{figure}
\vspace{-5.cm}
%\hspace{2.cm}
\epsfxsize=11.cm
\epsffile{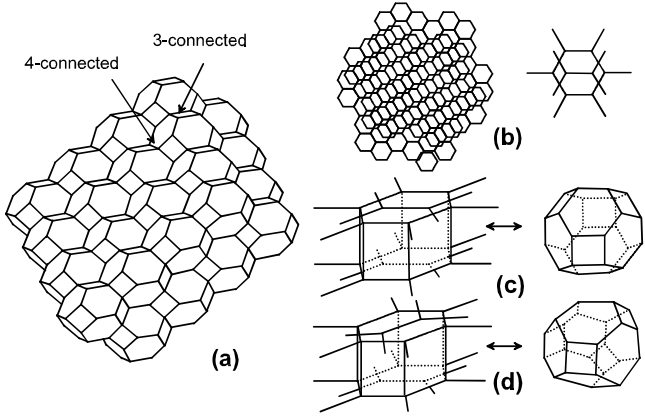}
\vspace{-5.5cm}
%\hspace{.5cm}
\caption{\footnotesize \label{f.Kelvin}
{\bf (a)} An example of 3D cellular system: the ``Kelvin froth''. 
A part of its shell-network is shown in {\bf (b)}, it is the 2D elliptic set of facets which are bounding externally 
a layer. 
{\bf (c)} Construction of the 3D ``Kelvin cell'' from the 2D shell network.
{\bf (d)} The ``twisted Kelvin cell''.
}
\vspace{0.cm}
\end{figure}

\vspace{0.cm}

In three dimensions, one can write a recursive equation similar to Eq.(\ref{2Drec}) \cite{AsBoRi,ShapeM}
\begin{equation}
K_{j+1} = s_j K_j - b_j K_{j-1} + c_j \;\;\;\;\;(\mbox{for $j \ge 1$}).
\label{3Drec}
\end{equation}
with
\begin{eqnarray}
s_j &=&{\displaystyle 1\over 2}\big( \langle f \rangle_j - 6  \big)\big( \langle n^N \rangle_j-4 \big) - b_j - 1 \nonumber \\
b_j &=& {\displaystyle \langle n^N \rangle_j-4 \over \langle n^N \rangle_{j-1}-4 } \nonumber \\
c_j &=& 2 \bigg( b_j -  s_j + 6 \big( \langle n^N \rangle_j-4 \big) +1 \bigg) \;\;\;\; .
\label{3Ds}
\end{eqnarray} 
Here $\langle f \rangle_j$ is the average number of neighbours of the cells in layer $j$. 
The quantity $\langle n^N \rangle_{j}$ is the average number of edges per face in the 2D elliptic 
set of facets which are bounding externally layer $j$ (the {\it shell-network}, see Fig.\ref{f.Kelvin}(a,b) 
and \cite{AsBoRi,ShapeM} for details).
The map starts with $K_0 = 0$ and $K_1 = f =$ {\it number of neighbours of the central cell}.
For example in the Kelvin structure one has $f = 14$, $\langle f \rangle_j = 14$ and 
$\langle n^N \rangle_{j}= 5 + (24 j^2 + 24 j +7)^{-1}$. 
By using the map (\ref{3Drec}) we obtain $K_2 = 50$, $K_3 = 110$ and in general $K_j = 12 j^2 +2$, 
which is the correct answer \cite{O'Keef:Book}.

Equations (\ref{2Drec}) and (\ref{3Drec}) take a much simpler form if the average topological 
properties of the cells in layer $j$ are independent on the layer number 
(i.e. $\langle n \rangle_j = \langle n \rangle$, $\langle f \rangle_j = \langle f \rangle$ and 
$\langle n^N \rangle_j=\langle n^N \rangle$). 
In general, these quantities can vary as one goes from one layer to the next (as for the Kelvin case). 
However, in the limit $j \rightarrow \infty$ they must converge towards the averages over the whole structure
(in the Kelvin case $\langle n^N \rangle_{j} \rightarrow 5$). 
Moreover, since the choice of the central cell is arbitrary, in disordered systems this asymptotic 
behaviour is reached much  faster if one averages over the central cell first.
Finally, the relations (\ref{2Drec}) and (\ref{3Drec}) are also valid when the ``central cell'' is a
cluster or even an infinite set of cells. 
For instance the starting configuration (the cells with $j=0$) can be an infinite linear strip of 
adjacent faces in 2D or a  planar layer of adjacent bubbles in 3D.
In this case the asymptotic behaviour and the average over the system can be set from the beginning. 

When the average topological properties of the cells in layer $j$ are independent on the layer number, 
Eqs.(\ref{2Drec}) and (\ref{3Drec}), give \cite{AsBoRi,RiUnp}
\begin{equation}
\left( \begin{array}{c}\tilde K_{j+1} \\ \tilde K_{j}  \end{array} \right) =
\left( \begin{array}{cc}s & -1 \\ 1 & 0\end{array} \right) \left( \begin{array}{c}\tilde K_{j} \\ 
\tilde K_{j-1}\end{array} \right) \;\;\;\; .
\label{Map}
\end{equation}
This equation is the logistic map \cite{Sch84}.
In two dimensions, $\tilde K_j = K_j$ is the number of polygonal cells which are making the layer $j$.
In three dimensions, for $s \not= 2 $, $\tilde K_j $ is the number of polyhedral cells making the layer $j$ plus the 
additive constant $c/(2-s)$. When $s = 2$  Eq.(\ref{Map}) cannot be used and Eq.(\ref{3Drec}) must be used instead.

\subsection{ Topological defects and non-SSI froths}
\begin{figure}
\vspace{-5.cm}
%\hspace{.5cm}
\epsfxsize=12.cm
\epsffile{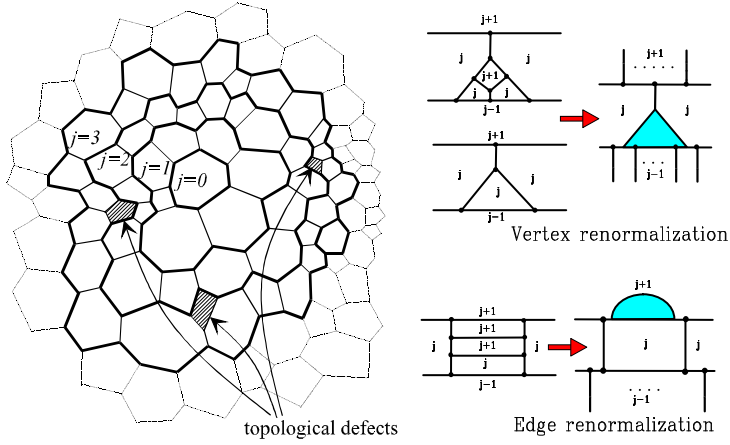}
\vspace{-6.cm}
%\hspace{-.5cm}
\caption{ \footnotesize \label{f.nonSSIfroth} \label{f.defects}
Example of non SSI froths (left). Defects, in layer $j$, are cells 
which have no neighbours in layer $j+1$. They are topological inclusions 
between the layers that can be considered as vertex- or edge-renormalizations (right). 
The topological map doesn't applies to such configurations,
but it applies on the shell-skeleton, which is the SSI structure obtained by eliminating the defects.}
\vspace{0.cm}
\end{figure}

\noindent
In almost all disordered froths, non-SSI inclusions (topological defects) are present. 
An example of non-SSI froth is given in Fig.\ref{f.nonSSIfroth}. 
A defect in layer $j$ is a cell (or a cluster of cells) which have no neighbours in layer $j+1$.
Examples of 2D topological defects are given in Fig.\ref{f.defects}.
The fraction ($\delta$) of defects in layer $j$ respect to the total number of cells,
tends to a constant as $j$ tends to infinity.
Typically, in 2D physical froths one finds $\delta \simeq 0.1$ \cite{AST}, whereas very disordered 
computer simulated froths have $0.15 < \delta < 0.6$ \cite{HABR}.

\vspace{0.cm}

Maps (\ref{2Drec}) and (\ref{3Drec}) apply to SSI froths only. 
But, for any froth, given a central cell, one can always single out the topological defects.
Then the reduction to the shell-skeleton is made by eliminating the defects.
The shell-skeleton is itself a space-filling froth hierarchically organized around the germ cell, 
and it is SSI by construction.
Maps (\ref{2Drec}) and (\ref{3Drec})  are therefore applicable on this structure.

\section{\label{s2} Space curvature from the map}  \label{S.MapCurv}
\begin{figure}
\vspace{-5.cm}
\epsfxsize=12.cm
\epsffile{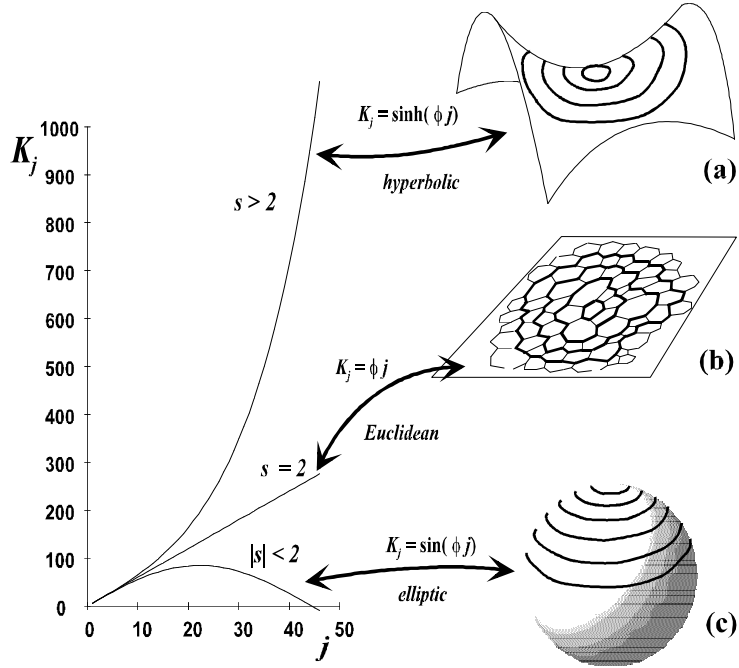}
\vspace{-4.cm}
\caption{\footnotesize \label{f.curvature}
The curvature of the space tiled by the froth is obtained from the behaviour of the number of cells ($K_j$) 
in successive layers as function of the topological distance ($j$) from the central 
cell.
The flat, Euclidean 2D space corresponds to a linear increment of $K_j$ with $j$ {\bf (b)}.
In elliptic spaces, $K_j$ first increases, reaches a maximum and then decreases {\bf (c)}.
In hyperbolic tilings this number increases  exponentially {\bf (a)}.
}
\vspace{0.cm}
\end{figure}

\noindent
Let start from a given cell of the froth and count its number of neighbours $K_j$ at a distance $j$.
We expect different behaviours for $K_j$ v.s. the distance $j$ depending on the curvature of the manifold
tiled by the froth, as schematically shown in Fig.\ref{f.curvature}.
Mathematically, these different behaviours can be easely studied from Eq.(\ref{Map}) when $s_j=s$  
independent of $j$.
In this case, one has the following solution of the map (\ref{Map}) \cite{AsBoRi},
\begin{equation}
\tilde K_j \propto \biggl( \exp(  \varphi j ) - \exp( -  \varphi j ) \biggr) \;\;\;\;.
\label{tra}
\end{equation}
The parameter $s$ separates the solutions into classes. 
The region $|s|<2$ is associated with $\varphi = i \cos^{-1}(s/2)$. 
It  has bounded, finite trajectories  in the $(j,K_j)$ parameter space. 
This is the characteristic behaviour of cellular tilings of elliptic manifolds (see Fig.(\ref{f.curvature}c)).
For $|s|>2$  one has $\varphi =  \cosh^{-1}(s/2)$, and unbounded exponential trajectories (typical of 
hyperbolic tilings, Fig.(\ref{f.curvature}a)). 
The point $s=2$ separates the elliptic from the hyperbolic regions. 
It corresponds to Euclidean tilings and the solution of Eqs.(\ref{2Drec}) and (\ref{3Drec}) is 
$K_j \propto j^{D-1}$, with $D$ the space-dimension (this is the behaviour expected from elementary geometry, 
Fig.(\ref{f.curvature}b)).
Note that this connection between the space curvature and the inflation parameter $s$ applies both in 
two and three dimensions.

\vspace{0.cm}

In two dimensions, equation (\ref{tra}) is the topological version of the relation (attributed to Gauss) 
between the circumference $C$ of a circle and its radius $\rho$ measured on a manifold of constant 
curvature ${{\cal R}}$ (where ${\cal R}$ = {\it Ricci curvature scalar} = {\it 2 $\times$ Gaussian curvature}),
\begin{equation}
C = { \pi \over \sqrt{-{{\cal R} \over 2}} }\biggl( 
\exp \bigl( \rho \sqrt{ -{{\cal R} \over 2}}\bigr)-
\exp\bigl(- \rho \sqrt{-{{\cal R} \over 2}}\bigr)\biggr)\;\;\;\; .
\label{C}
\end{equation}

In our case, $C$ represents the topological length ($K_j$) of the boundary of the cluster with topological 
radius $j$ and $\sqrt{-{{\cal R} \over 2}}$ is the parameter $\varphi$ of Eq.(\ref{tra}), given here in 
terms of the curvature.

Here are some examples of regular 2D froths with different $s$. 
To $s = -1$ corresponds an elliptic froth made with four triangles, i.e. the surface of a tetrahedron. 
The value $s = 0$ corresponds to an elliptic froth made with six squares, the surface of a cube. 
To $s = 1$ is associated an elliptic froth with pentagonal faces, the surface of a dodecahedron.
It is known that the hexagonal lattice ($s=2$) is the only regular froth which tiles the Euclidean plane. 
Froths with heptagonal ($s = 3$) or octagonal faces ($s = 4$) tile hyperbolic surfaces.
Two-dimensional disordered tilings can be generated by  Vorono\"{\i} \cite{Voronoi} tessellations around 
points at random on a surface with arbitrary curvature. 
Here the map must be studied experimentally case by case. 
In the Euclidean plane,  the linear behaviour of $K_j$ v.s. $j$ predicted by our map, has been experimentally 
confirmed by many measurements on natural and computer generated froths \cite{HABR,AST,Benoit}. 
The behaviour of $K_j$ in random Vorono\"{\i} networks constructed on curved surfaces is under investigation \cite{MMT}.

In two dimensions, the classification by the map (\ref{Map}) is identical to that provided by the combination 
of the Gauss-Bonnet theorem \cite{G-B} and Euler's equation:
\begin{equation}
{\int \!\!\!\!\! \int {{{\cal R} \over 2}} \, d {a} = 2 \pi \chi = {{\pi} \over 3} Q_T =
		{{\pi} \over 3}(6 - \langle n \rangle)N 
	= {{\pi} \over 3}(2 - s)N \;\;\;\; , }
\label{GB}
\end{equation}
where $N$ is the total number of cells in the system. 
Equation (\ref{GB}) shows clearly that the parameter $s < 2$ corresponds to global positive Gaussian curvature, 
$s>2$ to negative curvature and $s=2$ to the Euclidean limit.
In 2D, the determination of the curvature by the map (\ref{Map}) or by the Gauss-Bonnet theorem are identical, 
but the two methods are independent.
In three dimensions, the Euler's relation is homogeneous and from the Gauss-Bonnet formula it is not possible to 
distinguish the global curvature of a tiled manifold from the local average properties of its tiles. 
However, the map (\ref{Map}) is still applicable.
It provides therefore a general means of describing the curvature from topological considerations, also in 3D.

Here are two example of 3D froths with different inflation parameter $s$. 
The regular froth made by packing dodecahedras has $\langle f \rangle = 12$ and $\langle n^N \rangle = 5$, 
which substituted into Eq.(\ref{3Ds}) leads to $s=1$ , indicating therefore that this froth tiles a positively 
curved space. 
Indeed, it is known that it is a closed structure made with 120 dodecahedras.
The Kelvin froth of Fig.\ref{f.Kelvin} has $\langle f \rangle = 14$ and asymptotically 
$\langle n^N \rangle = 5$, corresponding to $s = 2$, which correctly indicate that the Kelvin's 
cells fill the Euclidean 3D space.

\section{\label{s3} Construction of three dimensional disordered structures from the map} \label{S.MapFree}
\noindent
Three-dimensional space-filling cellular systems are highly correlated structures. 
Each cell has a different shape and there exist in general very few configurations where all cells pack 
together and fill space without gaps or overlaps.
The space-filling condition is a very important constraint which strongly determines the properties 
of disordered structures.

\begin{figure}
\vspace{-3.5cm}
%\hspace{2.cm}
\epsfxsize=11.cm
\epsffile{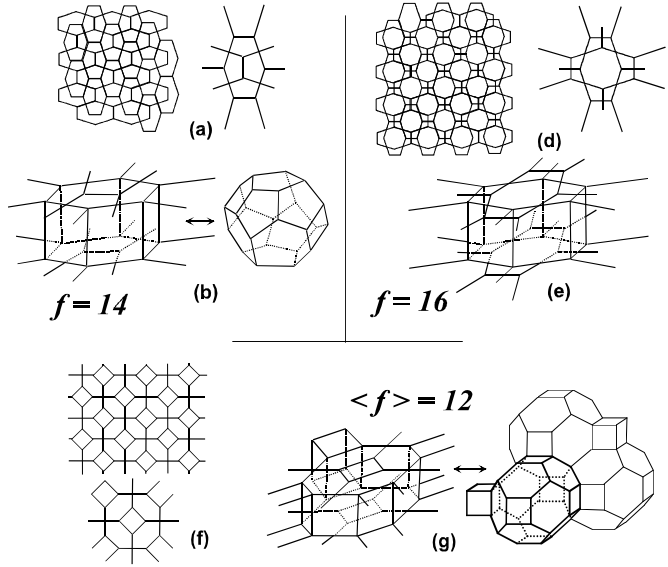}
\vspace{-4.5cm}
%\hspace{.5cm}
\caption{\footnotesize \label{f.3Dfroths}
Three dimensional froths generated by using the map starting from the shell network.
The shell network  is always the result of a superposition of
two three connected networks, in the examples {\bf (a)}, {\bf (d)} it is the superposition
of two hexagonal lattices.
}
\vspace{0.cm}
\end{figure}

In froths, the incidence numbers (number of edges on a vertex, of faces on a edge, etc.) 
are fixed at their minimal value. 
By contrast, the coordination number (number of faces of a 3D cell, number of edges of a 2D face) is a random 
variable and its average characterize the structure topologically \cite{AsRi}.
In two dimensions, the Euler relation fixes the average number of edges per face (see Eq.(\ref{GB})).
In three dimensions, the Euler relation associates the average number of edges per face $\langle n \rangle$ with 
the average number of faces per cell $\langle f \rangle$
\begin{equation}
\langle f \rangle={12 \over  6-\langle n \rangle} \;\;\;\; .
\label{f-n}
\end{equation}
So that the two average coordination numbers are related, and one of these two quantities is not constrained.

The representation of the cellular system in concentric layer around a given central germ cell can be also 
regarded as an iterative way of constructing disordered packings. 
In the Euclidean space (and in the asymptotic limit when  $\langle f \rangle_j$ and  $\langle n^N \rangle_j$ 
are independent of $j$), the inflation parameter of the map is fixed at $s=2$. 
For this value of $s$ Eq.(\ref{3Ds}) gives \cite{AsBoRi}
\begin{equation}
\langle f \rangle= 6 + {8 \over \langle n^N \rangle -4 } \;\;\;\; . 
\label{f-n1}
\end{equation}
This equation relates the average coordination ($\langle n^N \rangle$) of a two dimensional 
structure (the shell-network) with the average coordination ($\langle f \rangle$) of the three 
dimensional cellular system.
The shell-network is the result of a superposition of two elliptic 3-connected networks, the 
``incoming'' and the ``outgoing'' froths \cite{AsBoRi}. 
The pattern of edges constituting the shell-network sets the value of $\langle n^N \rangle$. 
Therefore, Eq.(\ref{f-n1}) allows us to construct systematically 3D Euclidean SSI froths starting from 2D 
shell-networks. 

The simplest 2D froth is the hexagonal lattice. 
The examples displayed in Fig.\ref{f.Kelvin} and Fig.\ref{f.3Dfroths}(a-e) illustrate the construction of ordered, 
monotiled 3D froths from a shell-network generated by different superpositions 
of two hexagonal lattices (see also \cite{AsBoRi,GLAZr}).
The cell in Fig.\ref{f.Kelvin}(c) is topologically equivalent to Kelvin's $\alpha$--tetrakaidecahedron 
(it builds up the Kelvin froth shown in Fig.\ref{f.Kelvin}(a)), and the cell \ref{f.Kelvin} (d), to its twisted 
variant \cite{Wil67,Wil68}. 
Both structures have $\langle f \rangle = 14$ and asymptotically $\langle n^N \rangle = 5$, 
and fill the Euclidean space.
In Fig.\ref{f.3Dfroths}(a) is shown part of a shell-networks with 5-sided faces, generated by the 
superposition of two ``squeezed'' hexagonal lattices and its unit cell (b) (see also \cite{AsBoRi,GLAZr}). 
This structure has again $\langle n^N \rangle = 5$, $\langle f \rangle = 14$ and is an Euclidean ($s=2$) space-filler. 
With another intersection of the two ``squeezed'' hexagonal lattices, one generates 
the shell-network shown in Fig.\ref{f.3Dfroths} (c), it has $\langle n^N \rangle = 4.8$ 
a 3D unit cell (\ref{f.3Dfroths} (d)) with $\langle f \rangle = 16$ (8 quadrilaterals, 6 
hexagons and 2 octagons). 
This unit cell is a monotile Euclidean space-filler that, as far as we know, was reported for the 
first time in \cite{AsBoRi}.
Recently, this tiling has been proposed \cite{O'Keeff16} as the structure of the ternary crystal $ThCr_3Si_4$.
Fig.\ref{f.3Dfroths} (e),(f) show an example of an Euclidean shell-structured-inflatable froth made of 
two different cells. 
The shell-network in Fig.\ref{f.3Dfroths} (e) has also two different tiles with 
$\langle n^N \rangle = 32/6 = 5.33...$ . 
The associated 3D unit cell (\ref{f.3Dfroths} (f)) has $\langle f \rangle = 12$ and 
$\langle n \rangle = 5$.
This structure is an Euclidean space-filler with $s=2$.
One can in general  show that any Euclidean shell-structured-inflatable froth made 
with topologically identical cells can be constructed from a shell-network generated 
by the superposition of two hexagonal lattices. 

\begin{figure}
\vspace{-4.5cm}
%\hspace{2.cm}
\epsfxsize=11.cm
\epsffile{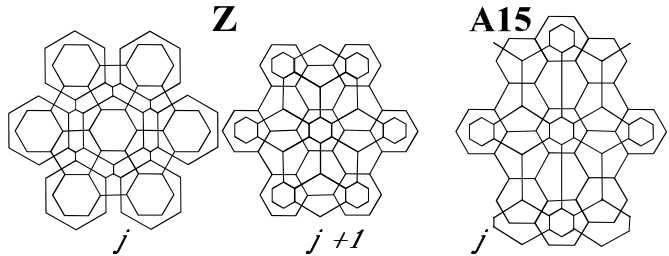}
\vspace{-7.5cm}
%\hspace{-1.cm}
\caption{\footnotesize \label{f.ZA15}
The shell network for the Z  and A15 phases (here the shells are planes).
In the Z and A15 shell networks there are respectively 12 and 14 pentagons per each hexagon.
Therefore, $\langle n^N \rangle$ result 66/12 and 76/15 for Z and A15 respectively.
In the Z phase, successive shells have alternating configurations $(j)$ and $(j+1)$. 
}
\vspace{0.cm}
\end{figure}
\vspace{0.cm}

We can restrict our attention to a special class of structures where the packed 
polyhedras have only pentagonal and hexagonal faces. 
In this case, the range of variability of $\langle f \rangle$ is restricted between 12 and 14 by 
Eq.(\ref{f-n}) and (\ref{f-n1}). 
The lower value corresponds to a structure with only pentagonal faces: a packing of dodecahedras, and it 
cannot be realized in Euclidean spaces (it is an elliptic system of 120 dodecahedras). 
Whether the upper value is realizable or not in Euclidean spaces is still an open question. 
(The only known structure with $\langle f \rangle=14$ made only with hexagon and pentagons is the Goldberg 
froth \cite{Goldberg}, but this packing is hyperbolic  \cite{AsBoRi}).
Two structures with coordinations number equal to 13.2  and 13.333... have been obtained by decurving 
iteratively the curved structure of 120 dodecahedras by introducing hexagonal faces \cite{SM85,RS88}.
A natural class of ordered structures made only with pentagons and hexagons are the 24 known TCP. 
They are Euclidean and have $13.333... \le \langle f \rangle \le 13.5$ \cite{FK58,FK59,ShoSho,Riv94,O'Keef}. 
The TCP A15, with $\langle f \rangle = 13.5$, is the known packing of polyhedral cells of equal volumes
with minimal interfacial area per cell \cite{WP}. 
In a previous paper \cite{AsBoRi}, we proved that all TCP structures could be described topologically by the 
map (\ref{Map}).
In figure \ref{f.ZA15} are given two examples of shell networks generated by combinations of pentagonal 
and hexagonal tiles only.
From these networks the structure of the Z phase and of the A15 phase can be constructed.
They have coordination $\langle n^N \rangle = 66/13 = 5.076...$ (Z) and  $\langle n^N \rangle = 76/15 = 5.066...$ (A15), that 
substituted into Eq.(\ref{f-n1}) give correctly $\langle f \rangle = 13.42...$ for Z and 
$\langle f \rangle = 13.5$ for A15.

\begin{figure}
\vspace{-3.cm}
%\hspace{1.5cm}
\epsfxsize=10.cm
\epsffile{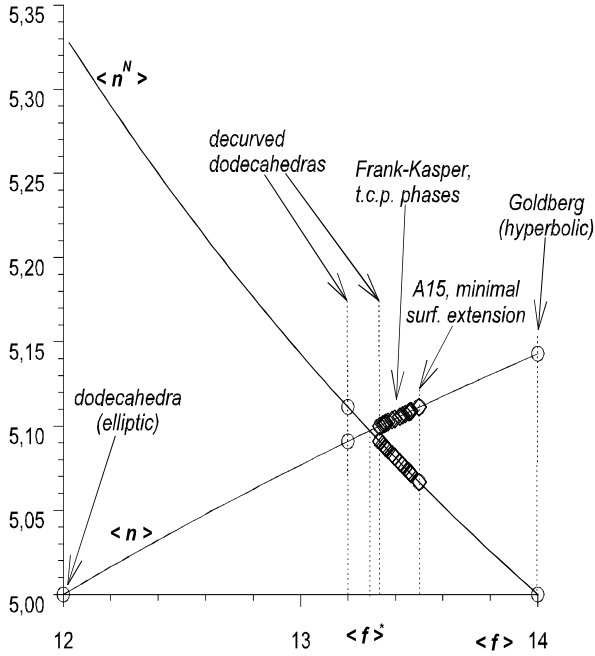}
\vspace{-3.cm}
%\hspace{-.5cm}
\caption{\footnotesize \label{f.numberOfNeigh}
Full lines: average number of edges per face in the whole froth ($\langle n \rangle$) 
and in the shell-network ($\langle n^N \rangle$) as function of the coordination number 
($\langle f \rangle$).
The circles and diamonds are some examples of known three dimensional cellular structures made 
by packing polyhedra with hexagonal and pentagonal faces only. Several of them (diamonds) occurs in 
nature (the TCP phases of metallic compounds).
}
\vspace{0.cm}
\end{figure}

\vspace{0.cm}

Figure \ref{f.numberOfNeigh} summarizes in a plot some of the previous considerations and results.
The two lines in Fig.(\ref{f.numberOfNeigh}) are plots of Eq.(\ref{f-n}) and (\ref{f-n1}). 
As one can clearly see, in a given structure, $\langle n \rangle$ and $\langle n^N \rangle$ are different in 
general. 
The packing is organized so that the faces of the shell-network have an average number of sides 
($\langle n^N \rangle$) different from that of the whole structure ($\langle n \rangle$).
But there is a point \cite{AsBoRi} where these two values coincide:
\begin{equation}
\langle n^N \rangle=\langle n \rangle=\langle n \rangle^*= {10 + 2\sqrt 7 \over 3}
\label{bestStrN}
\end{equation}
and, from (\ref{f-n}) or (\ref{f-n1})
\begin{equation}
\langle f \rangle^* = 8 + 2\sqrt 7 = 13.29...   \;\;\;\;.
\label{bestStrF}
\end{equation}

When $\langle f \rangle= \langle f \rangle^* = 13.29...$,  an arbitrary cell has the freedom to adhere to a 
layer by any subset of its faces, without adjustment. 
This freedom gives therefore many more local possibilities for building up the structure. 
A coordination number $\langle f \rangle = 13.29...$ characterizes therefore an hypothetical SSI topological 
packing which fills the Euclidean three-dimensional space with the largest possible number of local arrangements, 
that is with maximum entropy.
Coordination $13.29...$ lies in between the two froth obtained by decurving dodecahedras and it is inside the 
interval where $\langle f \rangle$ is ranging in natural structures and computer simulated froths  
\cite{Wil68,Oger96,FK58,FK59,ShoSho,O'Keef,Matzke,Hales}.

Note that, this structure has a coordination number which is irrational.
It cannot be therefore a periodic crystalline structure. It can only be approximated by disordered 
froths or quasi-crystalline structures.

\smallskip
\noindent
{\footnotesize The author acknowledge many discussions with N. Rivier. 
A special thanks to M. O'Keeffe, who sent us the preprints of his latests works. 
This work was partially supported by EU, HCM Program, ``FOAMPHYS'' network, 
contract ERBCHRXCT940542 and by TMR contract ERBFMBICT950380. }


\vspace{-0.3cm}
\begin{thebibliography}{}

%\baselineskip=10pt


\bibitem{AsBoRi}
\newblock{Aste, T., Boose, D.,  and Rivier, N. (1996) 
From one cell to the whole  froth: a dynamical map,
{\it Phys. Rev. E} {\bf 53} 6181-91.}


\bibitem{ShapeM}
\newblock{Aste, T., and  Rivier, N. (1997)
Topological Modeling of disordered cellular structures,
in {\it Shape Modeling and Applications} (IEEE Computer Society Press, Los Almitos) 2-9.}


\bibitem{dAT}
\newblock {Thompson, D'A.W., (1917, 1942) 
{\it On Growth and Form}, Cambridge Univ. Press., ch.7.}

\bibitem{WR84}
\newblock {Weaire, D., and Rivier, N. (1984), 
Soap; cells and statistics: Random patterns in two dimensions,
{\it Contemp. Physics} {\bf 25}, 59-99.}

\bibitem{AsRi}
\newblock{Aste, T., and   Rivier, N. (1995) 
Random cellular  froths  in spaces  of  any dimension and curvature, 
{\it J. Phys. A} {\bf 28}, 1381-98 }

\bibitem{Voronoi}
\newblock {Vorono\"{\i}, G.  (1908) 
Recherches sur les parall\'elo\`edres primitifs, 
{\it J. reine angew. Math.} {\bf  134}  198-287.}


\bibitem{CS}
\newblock {Charvolin, J.,  and Sadoc J.F. (1990) 
Structures built by amphiphiles and frustrated fluid films,  
{\it Colloque de Physique C} {\bf 7} 83-96.}

\bibitem{RD}
\newblock { Rivier, N., and Dubertret, B. (1995) 
Why does skin stay smooth?
The dynamics of tissues in statistical equilibrium, 
{\it Phil. Mag. B} {\bf 72} 311-322.}

\bibitem{SM85}
\newblock {Sadoc, J.F., and Mosseri, R.  (1985) 
Hierachical interlaced networks of disclination lines in non-periodic structures,
{\it J. Physique } {\bf 46} 1809-1826.}

\bibitem{SR}
\newblock { Sadoc J. F., and Rivier, N.  (1987)  
Hierarchy and disorder in non-crystalline structures,
{\it Phil. Mag. B }  {\bf 55 } 537-573.}

\bibitem{Sch84}
\newblock { Schuster, H. G. (1984) 
{\it Deterministic Chaos}  
(Physik-Verlag, Weinheim).}

\bibitem{RiUnp}
 Rivier, N. (1985) unpublished notes; (1986) seminar, Imperial College.

\bibitem{EulerPoincare}
\newblock{Henle, M.  (1979)
{\it A Combinatorial Introduction to Topology}, 
(Dover, New York).}

\bibitem{G-B}
\newblock {Kreyszig, E.  (1991) 
{\it Differential Geometry} 
(Dover, New York).}

\bibitem{Kelvin}
\newblock {Thomson, W. (Lord Kelvin) (1887) 
On the Division of Space  with Minimum Partitional Area,
{\it Phil. Mag. } {\bf 24}  503-514.}


\bibitem{TroUnp}
 Troadec, J.P. (1985) unpublished notes.

\bibitem{Fortes}
Fortes, M.A., and Pina, P.  (1993)
Average topological properties of successive neighbours of cells in random networks,
{\it Phil. Mag. B} {\bf 67}, 263-276. 


\bibitem{O'Keef:Book}
\newblock{O'Keeffe, M.  and Hyde, S. T.  (1996)
{\it Crystal Structures I: Patterns and Symmetry}, 
(Mineral. Soc. Amer., Washington D.C.).}


\bibitem{AST}
\newblock { Aste, T., Szeto, K. Y., and Tam, W. Y. (1996)
Statistical properties   and  shell  analysis  in  random   cellular structures,
{\it Phys. Rev. E} {\bf 54} 5482 -92.}

\bibitem{HABR}
\newblock{ Ohlenbusch, H. M., Aste, T., Dubertret, B., and Rivier, N.  (1997) 
The topological structure of 2D disordered cellular systems,
{\it European J. of Phys.}, submeted.}


\bibitem{Benoit}
\newblock{Dubertret, B.,  and Rivier, N.  (1997) 
The Renewal of the Epidermis: A topological Mechanism,
{\it Biophys. Journal} {\bf 73} 38-44.}

\bibitem{MMT}
\newblock {Demartino, M., Spagnuolo M., and Aste, T.,  work in progress.} 


\bibitem{GLAZr}
\newblock{Glazier, J. A., and Weaire, D. (1994)  
Construction of candidate minimal-area space-filling partitions,
{\it Phil. Mag. Lett.} {\bf 70}  351-356.} 


\bibitem{Wil67}
\newblock {Williams, W.M., (1967)
{\it Geometry, Structure, Enviroment},
(Southern Illinois University) 20.
}

\bibitem{Wil68}
\newblock {Williams, W.M., (1968)
{Space Filling Polyhedron: Its Relation to Aggregates of Soap Bubbles, Plant Cells, 
and Metal Crystallites},
{\it Science} {\bf 161} 276-277..
}
\bibitem{O'Keeff16}
\newblock{O'Keeffe, M., (1997)
On a space filling polyhedron of Aste, Boos\'e and Rivier,
{\it Phil. Mag. Lett.}, to appear.}

\bibitem{Goldberg}
\newblock {Goldberg, M., (1934) 
{\it Tohoku Math. J.} {\bf 40} 226.}

\bibitem{RS88}
\newblock { Rivier N., and Sadoc, J. F.   (1988)
The Local Geometry of Disorder
{\it J. non-cryst. Solids} {\bf 106} 282-85.}

\bibitem{Oger96}
\newblock {Oger, L.,Gervois, A., Troadec, J. P.,  and Rivier, N.   (1996)
Voronoi tessellation of packings of spheres. Topological
correlations and statistics,
{\it Phil. Mag. B} {\bf 74} 177-197.}


\bibitem{FK58}
\newblock {Frank, F. C.,  and Kasper, J. S. (1958) 
Complex Alloy Structures Regarded as Sphere Packings. I Definitions and Basic Principles, 
{\it  Acta crystallogr. } {\bf 11 }  184-190. }

\bibitem{FK59}
\newblock {Frank, F. C.,  and Kasper, J. S. (1959) 
Complex Alloy Structures Regarded as Sphere Packings. 
II Analysis and Classification of Representative Structures,
{\it  Acta crystallogr. } {\bf 12 }  483-499. }

\bibitem{ShoSho}
\newblock {Shoemaker D. P.,  and Shoemaker, C. B.  (1986)
Concerning the Relative Numbers of Atomic Coordination Types in Tetrahedrally 
Close Packed Metal Structures,
{\it Acta crystallogr. B}  {\bf 42 }  3-11.}


\bibitem{Riv94}
\newblock { Rivier, N. (1994)
Kelvin's conjecture on minimal froths and the counter-example of Weaire and Phelan,
{\it Phil. Mag. Lett.} {\bf 69}  297-303}

\bibitem{O'Keef}
\newblock{M. O'Keeffe, (1997)
Sphere Packings and Space Filling by Congruent Simple Polyhedra,
{\it Acta. Cryst.} submetted.}

\bibitem{WP}
\newblock {Weaire, D. and Phelan, R.  (1994)
A counter-Example to Kelvin's Conjecture on Minimal Surfaces,
{\it Phil. Mag. Lett.} {\bf 69} 107-110.}


\bibitem{Matzke}
\newblock {Matzke, E.B.  (1946) 
The Three-Dimensional Shape of Bubbles in Foam,
{\it Am. J. Botany} {\bf 33} 58-80.}

\bibitem{Hales}
\newblock {Hales, S.  (1727) 
{\it Vegetable Staticks}, 
(Innys and Woodward, London).}





%\bibitem{RA96}
%\newblock{ N. Rivier and  T. Aste, in {\it The geometry of curved Surfaces and Interfaces},
%ed. by J. Klinowski A. L. Makay, (The Philosophical Transaction A of the Royal Society, London 1996), {\bf 354} 2055-69.}

%\bibitem{Alexander}
%\newblock {J. W. Alexander, {\it Ann. Math.} {\bf 31} (1930) 292.}



%\bibitem{RLiss82}
%\newblock {N. Rivier and A. Lissowski, {\it J. Phys. A } {\bf 15} (1982) L143.}

%\bibitem{PSR91}
%\newblock {M.A. Peshkin, K.J. Stradburg and N. Rivier, {\it  Phys. Rev. Lett.} {\bf 67} (1991) 1803.}

%\bibitem{RinNadal}
%\newblock {N. Rivier, in {\it From Statistical Physics to Statistical Inference and Back}, edited by P. Grassberger and J.-P. Nadal (Kluwer, Dordrecht 1994) 77-93.}



%\bibitem{RinForma}
%\newblock{ N. Rivier,  {\it FORMA} {\bf 11} (1996) 195-98.}



%\bibitem{Amorph}
%\newblock {T. Aste and N. Rivier, work in progress.}


%\bibitem{Euler}
%\newblock{Euler ????}

%\bibitem{ItzDru88}
%\newblock {C. Itzykson and J.M. Drouffe, {\it Statistical Field Theory} (Cambridge University Press 1988) vol.2, chap.11 .}

%\bibitem{Mei53}
%\newblock {J.L. Meijering, {\it Philips Res. Rep.} {\bf 8} (1953) 270.}

%\bibitem{Riv82}
%\newblock {N. Rivier, {\it J. Physique (Coll.)} {\bf 43} (1982) C9-91.}



\end{thebibliography}
\end{document}